==============================
\documentstyle{mn}
\newif\ifAMStwofonts

\def\gsim{\;\lower4pt\hbox{${\buildrel\displaystyle >\over\sim}$}\;}
\def\lsim{\;\lower4pt\hbox{${\buildrel\displaystyle <\over\sim}$}\;}
\def\grls{\;\lower4pt\hbox{${\buildrel\displaystyle >\over <}$}\;}
\ifoldfss
  \ifCUPmtlplainloaded \else
    \NewTextAlphabet{textbfit} {cmbxti10} {}
    \NewTextAlphabet{textbfss} {cmssbx10} {}
    \NewMathAlphabet{mathbfit} {cmbxti10} {} 
    \NewMathAlphabet{mathbfss} {cmssbx10} {} 
  \fi
  \ifAMStwofonts
    \ifCUPmtlplainloaded \else
      \NewSymbolFont{upmath} {eurm10}
      \NewSymbolFont{AMSa} {msam10}
      \NewMathSymbol{\upi}     {0}{upmath}{19}
      \NewMathSymbol{\umu}     {0}{upmath}{16}
      \NewMathSymbol{\upartial}{0}{upmath}{40}
      \NewMathSymbol{\leqslant}{3}{AMSa}{36}
      \NewMathSymbol{\geqslant}{3}{AMSa}{3E}

    \fi
  \fi
\fi 

\ifnfssone
  \newmathalphabet{\mathit}
  \addtoversion{normal}{\mathit}{cmr}{m}{it}
  \addtoversion{bold}{\mathit}{cmr}{bx}{it}
  \newmathalphabet{\mathbfit} 
  \addtoversion{normal}{\mathbfit}{cmr}{bx}{it}
  \addtoversion{bold}{\mathbfit}{cmr}{bx}{it}
  \newmathalphabet{\mathbfss} 
  \addtoversion{normal}{\mathbfss}{cmss}{bx}{n}
  \addtoversion{bold}{\mathbfss}{cmss}{bx}{n}
  \ifAMStwofonts
    \ifCUPmtlplainloaded \else
      \UseAMStwoboldmath
      \makeatletter
      \new@mathgroup\upmath@group
      \define@mathgroup\mv@normal\upmath@group{eur}{m}{n}
      \define@mathgroup\mv@bold\upmath@group{eur}{b}{n}
      \edef\UPM{\hexnumber\upmath@group}
      \new@mathgroup\amsa@group
      \define@mathgroup\mv@normal\amsa@group{msa}{m}{n}
      \define@mathgroup\mv@bold\amsa@group{msa}{m}{n}
      \edef\AMSa{\hexnumber\amsa@group}
      \makeatother
      \mathchardef\upi="0\UPM19
      \mathchardef\umu="0\UPM16
      \mathchardef\upartial="0\UPM40
      \mathchardef\leqslant="3\AMSa36
      \mathchardef\geqslant="3\AMSa3E
    \fi
  \fi
\fi 

\ifnfsstwo
  \DeclareMathAlphabet{\mathbfit}{OT1}{cmr}{bx}{it}
  \SetMathAlphabet\mathbfit{bold}{OT1}{cmr}{bx}{it}
  \DeclareMathAlphabet{\mathbfss}{OT1}{cmss}{bx}{n}
  \SetMathAlphabet\mathbfss{bold}{OT1}{cmss}{bx}{n}
  \ifAMStwofonts
    \ifCUPmtlplainloaded \else
      \DeclareSymbolFont{UPM}{U}{eur}{m}{n}
      \SetSymbolFont{UPM}{bold}{U}{eur}{b}{n}
      \DeclareSymbolFont{AMSa}{U}{msa}{m}{n}
      \DeclareMathSymbol{\upi}{0}{UPM}{"19}
      \DeclareMathSymbol{\umu}{0}{UPM}{"16}
      \DeclareMathSymbol{\upartial}{0}{UPM}{"40}
      \DeclareMathSymbol{\leqslant}{3}{AMSa}{"36}
      \DeclareMathSymbol{\geqslant}{3}{AMSa}{"3E}
    \fi
  \fi
\fi 

\ifCUPmtlplainloaded \else
  \ifAMStwofonts \else 
    \def\upi{\pi}
    \def\umu{\mu}
    \def\upartial{\partial}
  \fi
\fi

\title[Jupiter's QP-40 Polar Burst Activities]
{On the Importance of Searching for \\
Oscillations of the Jovian Inner Radiation\\
Belt with a Quasi-Period of 40 Minutes}

\author[Lou and Zheng]
{
Yu-Qing Lou$^{!\ \dagger\ *}$
and Chen Zheng$^{\dagger}$\\
$^{\dagger}$Physics Department, Tsinghua Center for Astrophysics, 
Tsinghua University, Beijing 100084, China\\ 
$^{!}$National Astronomical Observatories,
Chinese Academy of Sciences, A20, Datun Road, 
Beijing, 100012 China\\  
$^{*}$Department of Astronomy and Astrophysics, 
The University of Chicago,
Chicago, Illinois 60637 USA  }
\date{Accepted ....
      Received ...;
      in original form ...}
\pagerange{\pageref{firstpage}--\pageref{lastpage}}

\pubyear{2003}

\begin{document}

\maketitle

\label{firstpage}

\begin{abstract}
Experiments aboard the Ulysses spacecraft discovered quasi-periodic 
bursts of relativistic electrons and of radio emissions with 
$\sim 40-$minute period (QP-40) from the south pole of Jupiter 
in February 1992. Such polar QP-40 burst activities were found to 
correlate well with arrivals of high-speed solar winds at Jupiter.
We advance the physical scenario that the inner radiation belt (IRB) 
within $\sim 2-3$ Jupiter's radius $R_J$, where 
relativistic electrons are known to be trapped via synchrotron 
emissions, can execute global QP-40 magnetoinertial oscillations 
excited by arrivals of high-speed solar winds. Modulated by such 
QP-40 IRB oscillations, relativistic electrons trapped in the IRB 
may escape from the magnetic circumpolar regions during a certain 
phase of each 40-min period to form circumpolar QP-40 electron 
bursts. Highly beamed synchrotron emissions from such QP-40 burst 
electrons with small pitch angles relative to Jovian magnetic fields 
at $\sim 30-40R_J$ give rise to QP-40 radio bursts with typical 
frequencies $\lsim 0.2$MHz. We predict that the synchrotron 
brightness of the IRB should vary on QP-40 timescales upon arrivals 
of high-speed solar winds with estimated magnitudes $\gsim 0.1$Jy, 
detectable by ground-based radio telescopes. The recent discovery 
of $\sim 45$-min pulsations of Jupiter's polar X-ray hot spot by 
the High-Resolution Camera (HRC) of the Chandra spacecraft provides 
a strong supporting circumstantial evidence that the IRB neighborhood 
did oscillate with QP-40 timescales. Using the real-time solar wind 
data from the spacecraft Advanced Composition Explorer (ACE), we show 
here that such QP-40 pulsations of Jupiter's polar X-ray 
hot spot did in fact coincide with the arrival of high-speed solar wind 
at Jupiter. We note also that a properly sampled data of simultaneous 
far-ultraviolet images of auroral ovals obtained by the Hubble Space 
Telescope imaging spectrograph (HST-STIS) would have contained QP-40 
oscillatory signatures. By our theoretical analysis, we offer several 
predictions that can be tested by further observations.
\end{abstract}


\section{Introduction}

Quasi-periodic bursts of relativistic electrons 
(Simpson et al. 1992; McKibben et al.
1993; Desch 1994) and of accompanied radio emissions (MacDowall et al. 
1993) were discovered by the Ulysses spacecraft a decade ago. The burst 
electron energy ranges from a few to $\gsim 10$ MeV and the frequency 
of radio bursts is usually $\lsim 200$ kHz with occasional rises to 
$\sim 700$ kHz. These burst activities, characterized by a quasi-period 
of 40 minutes (QP-40), were inferred to occur in the south polar 
direction of Jupiter and were found to closely correlate with arrivals 
of high-speed solar winds at Jupiter (see figs. 11 and 12 of MacDowall 
et al. 1993; Bame et al. 1992). Polarizations of these radio bursts 
are predominantly right-handed. There were other evidence, though less 
certain, for Jovian QP-40 phenomena such as magnetic field fluctuations 
and proton fluxes from observations of various spacecraft 
(e.g., Balogh et al. 1992; Schardt et al.
1981).

Based on a theoretical magnetohydrodynamic (MHD) analysis, we 
proposed (Lou 2001) that the underlying physical mechanism 
may involve global QP-40 magnetoinertial oscillations of 
Jupiter's inner radiation belt (IRB) within $\sim 2-3R_J$. 
In the scenario of QP-40 IRB oscillations, several seemingly 
disparate key phenomena can be plausibly linked together. 
%
%
Variabilities from weeks to years of the Jovian system have been 
searched for in response to solar wind variations (e.g., Bolton 
et al. 1989; Miyoshi et al. 1999). The central question posed here 
is whether the synchrotron brightness of the IRB varies on QP-40 
timescales and shorter ones resulting from solar wind speed 
variations at Jupiter. 

The recent Chandra observations (Gladstone et al. 2002) detected 
QP-40 pulsations of north polar X-ray hot spots on Jupiter. 
From space, Cassini synchrotron radio observations at 2.2cm 
(Bolton et al. 2002) revealed relativistic electrons with energies 
$\gsim 50$ MeV inside the IRB. These observations provide a 
circumstantial evidence for QP-40 IRB oscillations. In this Letter, 
we derive physical consequences from empirical facts, provide 
relevant estimates, propose further observational tests, and 
establish that $\sim 45-$min pulsations of Jupiter's polar X-ray 
hot spot did coincide with an arrival of high-speed solar wind at 
Jupiter using the data from the spacecraft Advanced Composition 
Explorer (ACE). It would be significant to actively search for 
QP-40 IRB synchrotron brightness variations upon arrivals of 
high-speed solar winds at Jupiter using ground-based radio 
telescope facilities. Once established, this would enable us 
to study the dynamics of the IRB both observationally and 
theoretically from a new perspective.

\section{The Physical Scenario}

By tracing QP-40 radio burst directions (MacDowall et al. 1993) 
in the plane of sky, comparing arrival times of various 
energetic particle species (Zhang et al. 1995), and analyzing 
electron anisotropies (Zhang et al. 1993), it is fairly certain 
that these charged particles bursted from Jupiter's south 
magnetic polar region. Jupiter has long been known to be an 
important source of relativistic electrons populating the 
heliosphere (Simpson et al. 1975; Nishida 1976). While specific 
aspects remain to be understood, the Jovian magnetospheric system 
interacting with the solar wind  appears to be capable of producing 
relativistic electrons to supply the IRB and to compensate magnetic 
polar leaks.

What is then the origin or source of such polar QP-40 bursts 
of relativistic electrons and ions? By all accounts, it is 
plausible (Lou 2001) that QP-40 burst electrons and ions leak out 
from the narrow circumpolar zone separating the magnetic polar
region with open magnetic fields from the adjacent Jovian IRB with 
closed magnetic fields. 
The IRB can trap relativistic electrons with energies up to 
$\sim 50$MeV or higher, as inferred from IRB synchrotron emissions 
with wavelengths of $\sim 2.2-90$ cm (Bolton et al. 2002; de Pater
1984; Roberts et al. 1984; Sault et al. 1997).
Moreover, by combined effects of Jupiter's fast rotation with a
$\sim 10$-hour period and a strong dipole magnetic field with 
a polar surface strength $|\vec B|$ of $\sim 10-14.4$G, the 
IRB is capable of magnetoinertial oscillations with periods of 
$\sim 40-50$ minutes and shorter ones (Lou 2001). 

Such magnetoinertial oscillations of a rotating IRB involve both 
Lorentz and Coriolis forces. For low-frequency and large-scale 
oscillations, the mode frequency is hybrid of Alfv\'en and rotation 
frequencies; for high-frequency and small-scale oscillations, the 
mode frequencies are essentially those of globally trapped fast MHD 
waves. For the hybrid mode of the lowest frequency, the IRB plasma 
swings to and fro about the rotation axis.

With this scenario in mind, the gross overall correlations among 
QP-40 bursts of radio emissions (MacDowall et al. 1993), of 
relativistic electrons (Simpson et al. 1992; McKibben et al. 1993), 
of protons, and occasionally, of alphas (Zhang et al. 1995) seems 
to indicate that
(i) QP-40 polar electron bursts are most likely
responsible for QP-40 polar radio bursts, and
(ii) a global resonant oscillatory mechanism may underlie 
the quasi-periodicity of $\sim 40$ min for polar bursts 
of relativistic electrons and ions (Schardt et al. 1981; 
Zhang et al. 1995).

To relate QP-40 IRB oscillations and QP-40 circumpolar bursts of 
electrons, we invoke large-scale asymmetries as well as small-scale 
irregulaties in magnetic field structures of the IRB (Lou 2001). 
During a certain phase of each IRB pulsation period, relativistic 
electrons may drift across thin vulnerable layers randomly 
spreading along circumpolar magnetic footpoints with a 
perpendicular gradient drift speed 
$\dot{\vec {\cal R}_{\perp}}=-m_e\gamma v_{\perp}^2c\vec B
\times\nabla |\vec B|/(2eB^3)$ into a narrow circumpolar strip and thus
give rise to an electron burst, where $m_e$ is the electron mass, 
$\gamma$ is the Lorentz factor, $v_{\perp}$ is the electron velocity
perpendicular to $\vec B$, $c$ is the speed of light, and $e$ is the
electron charge. Hence, QP-40 IRB oscillations lead to QP-40 
bursts of relativistic electrons and ions from magnetic circumpolar
zones. QP-40 radio bursts are then produced by highly beamed
synchrotron radio emissions from such escaped relativistic 
electrons with very small pitch angles $\alpha$ (e.g., 
$\alpha\sim 6-4\times 10^{-3}$ at $\sim 30-40R_J$) relative 
to magnetic field lines such that radio burst frequencies are 
typically $\lsim 200$ kHz detected by Ulysses (MacDowall 2001). 
It requires relativistic electrons of higher $\gamma$ to produce 
occasional rising frequencies of up to $\sim 700$ kHz at Ulysses.
That onsets of QP-40 radio bursts (McKibben et al. 1993; Desch 1994) 
sometimes precede relevant electron bursts by $\sim 4-8$ min are 
likely caused by radio emissions from those relativistic electrons 
that travel along neighboring magnetic field lines but that are not 
intercepted by Ulysses (Lou 2001). Admittedly, we do not yet know 
details of circumpolar electron leak processes by lacking clues 
of magnetic field inhomogeneities and irregularities or defects. 

\section{Theoretical Considerations}

In reference to observations, we now provide 
theoretical analyses and corresponding predictions.

\subsection{Excitations of Global IRB Oscillations}

Empirically, onsets of enhanced QP-40 burst activities correlate 
well with arrivals of high-speed solar winds at Jupiter (MacDowall 
et al. 1993). While the solar wind mass flux remains roughly 
constant for either fast or slow winds, variations in the wind 
speed $U$ ($\sim 800-400{\hbox{ km s}^{-1}}$) cause the radial 
size of the sunward magnetosphere to change drastically 
($\sim 50-100R_J$), where $R_J\cong 7.14\times 10^9$ cm.
This offers a valuable clue for the 
excitation of QP-40 IRB oscillations and thus for the observed 
correlation of QP-40 polar burst activities with high-speed 
solar winds. Drastic changes of solar wind speed at the Jovian 
magnetosphere or persistent Jovian magnetospheric compressions 
sustained during a high-speed solar wind phase with irregular 
intermittent relaxations caused by wind speed variations can 
both resonantly drive QP-40 magnetoinertial IRB pulsations and 
induce neighboring magnetic field oscillations (Balogh et al. 
1992)
through the conservation of the magnetospheric angular 
momentum (Nishida \& Watanabe 1981). Such MHD pulsations of 
the IRB should then lead to QP-40 brightness variations.

\subsection{Estimates for IRB Brightness Variations}

The solar wind mass flux is estimated by
$4\pi\rho U D_J^2$,
where $\rho$ is the wind mass density and $D_J$ is 
Jupiter's distance to the Sun. Changing from slow to 
fast winds at Jupiter,
the wind ram pressure $\rho U^2$ increases by a factor of $\sim 2$.
After a transient time of adjustment and for a negligible IRB 
thermal pressure, this increase of wind ram pressure is grossly
balanced by an increase of magnetic pressure $B^2/(8\pi)$ of 
the IRB temporarily. Thus, the relative field strength variation 
$\delta B/B$ in the IRB may be as large as $\sim 40\%$ stirred 
by drastic solar wind speed variations; and by the magnetic flux 
conservation, the radial extent of the IRB may vary by 
$\sim 20\%$ accordingly. 

In the IRB, for a power-law distribution of electron number 
density $N(\gamma)\propto\gamma^{-S}d\gamma$ in the energy interval
$(\gamma, \gamma+d\gamma)$, the spectral intensity $I_{\nu}$ of 
radiation is $I_{\nu}\propto B^{(S+1)/2}\nu^{-(S-1)/2}$
(Ginzburg \& Syrovatskii 1965; Rybicki \& Lightman 1979). At a given 
frequency $\nu$, one has $\delta I_{\nu}/I_{\nu}\cong (S+1)\delta B/(2B)$. 
For the Jovian IRB, the spectral index $S>1$ and $I_{\nu}$ takes on 
values of $\sim 0.44\pm 0.15$, $\sim 3$, $\sim 4.02\pm 0.08$, and 
$\sim 5.15\pm 0.7$Jy at $\nu=13.8$, 5, 2.3, and 0.333 GHz, respectively 
(Bolton et al. 2002). Thus, in two separate frequency ranges $0.333-5$GHz 
and $5-13.8$GHz, $(S-1)/2\sim 0.19$ and $\sim 1.9$, respectively. 
Conservatively, $\delta I_{\nu}$ is estimated to be $\gsim 0.1$Jy. As 
a crucial test, such QP-40 IRB brightness variations in wavelengths 
$\sim 6-90$ cm should be searched for using ground-based radio telescopes. 

\subsection{Polarization Properties of Radio Bursts}

There are several physical reasons and observational tests 
to support and to further check our scenario. The radiation 
electric field $\vec E_{rad}$ from an accelerating electric 
charge $q$ is given by (e.g., Rybicki \& Lightman 1979)
$$
\vec E_{rad}(\vec r, t)={q\over c}\bigg[{\vec n\over\kappa^3R}
\times\{(\vec n-\vec\beta)\times \dot{\vec\beta\}}\bigg]\ ,
\eqno(1)
$$
where brackets denote variables at retarded times,
$\vec\beta\equiv \vec u/c$ is the particle velocity normalized by
$c$, $\dot{\vec\beta}$ is the time derivative of $\vec\beta$,
$\vec n$ is the unit vector along the line of sight distance $R$ 
at retarded times, and $\kappa\equiv 1-\vec n\cdot\vec\beta$.
The radiation magnetic field is $\vec B_{rad}(\vec r, t)\equiv 
[\vec n\times\vec E_{rad}(\vec r, t)]$. The south polar magnetic 
field $\vec B$ points towards the Jupiter. For a relativistic 
electron streaming outward from the south pole with a small pitch 
angle $\alpha$ (nearly anti-parallel to $\vec B$), the instantaneous 
$\vec E_{rad}$ is nearly along $\dot{\vec\beta}$. Given an 
electron's right-hand gyration with respect to $\vec B$, the radio 
polarization at Ulysses should be right-handed; this qualitative 
conclusion has also been confirmed by more detailed numerical 
computations. This right-handed radio polarization should prevail 
for a bunch of electrons so long as their spatial distribution is 
not completely random. This theoretical result is consistent with 
the Ulysses Radio and Plasma Wave Experiment (URAP) observations 
(MacDowall 1993).

In year 2004, Ulysses will have a second rendezvous with Jupiter
in the northern heliosphere with a closest approach of $\sim 1000R_J$.
For global IRB oscillations and qualitatively similar polar
$\vec B$ configurations as well as level of irregularities, QP-40 
bursts of relativistic electrons and of accompanied radio emissions 
from the north pole are anticipated, especially during arrivals of 
high-speed solar winds at Jupiter. While Ulysses particle instruments 
cannot intercept QP-40 bursts of relativistic 
electrons (one still expects to observe a gradual increase of 
relativistic electron flux towards Jupiter), URAP will have a good 
opportunity to observe north polar QP-40 radio bursts outside the 
Jovian magnetosphere. As $\vec B$ points outward from the Jovian 
north pole, we predict that polarizations of north circumpolar 
QP-40 radio bursts should be predominantly left-handed.

\subsection{Frequencies of QP-40 Radio Bursts}

Let us now estimate typical frequency components of a radio burst. 
As relativistic electrons leak out quasi-periodically during a 
certain phase in each period of IRB QP-40 oscillations, radio 
burst emissions are produced by the synchrotron process from 
relativistic electrons in nearly anti-parallel motions outward 
along south polar magnetic field lines. Such IRB
relativistic electrons initially drift across the narrow 
circumpolar zone about the magnetic axis and they gyrate 
transverse to the local surface $\vec B$, radiating intensely
in a perpendicular plane due to the relativistic beaming effect 
(i.e., strongest emissions within an angle $\lsim 1/\gamma$ 
about $\vec v$ direction).
The total power emitted by an electron is
$P(\gamma)=2(\gamma^2-1)e^4B^2\sin^2\alpha/(3m_e^2c^3)$. 

As electrons
stream outward from circumpolar magnetic footpoints with gyroradii
$r_c\sim\gamma m_ec^2/(eB)=1.7\times 10^3\gamma B^{-1}$ cm $\ll R_J$,
the magnetic mirror force rapidly converts transverse gyrations to 
parallel motions by conserving both particle energy 
${\cal E}\equiv\gamma m_ec^2$ and magnetic moment
${\cal E}_{\perp}/|\vec B|$ where ${\cal E}_{\perp}$ is the kinetic
energy of transverse gyration. The synchrotron emission cone ahead 
of a relativistic electron gyrating around $\vec B$ shrink in conal 
angle towards the forward direction of $\vec v$ with decreasing
characteristic frequencies (Ginzburg \& Syrovatskii 1965).

For a dipole magnetic field $|\vec B|\propto r^{-3}$ and a polar surface 
$|\vec B|$ strength of $\sim 10$G, the pitch angle $\alpha$ of electrons, 
independent of $\gamma$, becomes $\sim 6\times 10^{-3}$ at $\sim 30R_J$ 
or $\sim 4\times 10^{-3}$ at $\sim 40R_J$. The characteristic synchrotron 
emission frequency $\nu_c$ from a gyrating electron is given by
$
\nu_c=0.29\times 3\gamma^2eB\sin\alpha/(4\pi m_ec) 
$
(Rybicki \& Lightman 1979). For parameters of interest, 
we have $\nu_c\sim 3\gamma^2$ and $\sim 0.87\gamma^2$ at 
$\sim 30R_J$ and $\sim 40R_J$, respectively. For typical 
spectral profiles of QP-40 radio bursts, the dominant 
peak falls between $\sim 10-80$kHz (MacDowall et al. 1993). 
\footnote{A reduced intensity at about $40-50$ kHz may not 
be a common feature of the QP-40 bursts. The transition
from the URAP RAR Low to High band receivers occurs at
$\sim 50$ kHz.}
For a power-law number distribution
of IRB electrons in $\gamma$, this would imply 
relativistic electrons with a $\gamma$ range of as high as 
$\sim 160-300$. The recent Cassini synchrotron observation in space at 
2.2cm (13.8GHz) revealed electron energies of $\sim 50$MeV inside the 
IRB, with further hints of a high-energy tail with $\gamma\gsim 200$ 
at $r\gsim 2R_J$ (Bolton et al. 2002). 
%
If $\sin\alpha$ is taken to be $\sim 1$ 
as commonly assumed, then $\nu_c$ would be 
much higher than several hundred kilohertz.

\subsection{QP-40 Pulsations of Polar X-Ray Hot Spots}

With both magnetic poles being qualiatively equal, we focus
on pulsations of northern auroral X-ray hot spot of Jupiter 
with a $\sim 45$-minute period discovered lately (Gladstone 
et al. 2002) using the high-resolution camera (HRC) of the 
Chandra X-ray Observatory on 18 December 2000.
While physical processes leading to such polar X-ray hot spots 
inside the main far-ultraviolet (UV) polar auroral oval are 
currently unexplained, their QP-40 pulsations nontheless 
offer extremely valuable diagnostics for probing the inner 
magnetospheric environment. 

In the scenario of QP-40 IRB oscillations (Lou 2001), the
northern auroral X-ray hot spots should pulsate with a quasi-period 
of $\sim 40$ min as QP-40 magnetoinertial oscillations of the IRB
will affect, through fast MHD wave transmissions, the immediate 
environs that include circumpolar zones of open magnetic fields, and 
leave QP-40 oscillatory signatures there (Balogh et al. 1992). Almost 
certainly, $\sim 45$-min pulsations of southern auroral X-ray hot 
regions were not seen this time (Gladstone et al. 2002) primarily 
owing to an 
unfavorable viewing geometry from the Chandra HRC. We anticipate 
that, similar to QP-40 polar burst activities and IRB oscillations 
(MacDowall et al. 1993), pulsation magnitudes of such X-ray hot 
spots poleward of the far-UV auroral oval should be also enhanced 
upon arrivals of high-speed solar winds at Jupiter. One primary goal 
of the Chandra and HST campaigns supporting the Cassini fly-by is 
to search for connections between Jovian auroral transients and the 
interaction of the solar wind with Jupiter's magnetosphere; in fact, 
this is already in hand but not yet fully appreciated. 

It is promising to pursue a direct detection of Jupiter's 
QP-40 IRB brightness variations in the wavelength range 
of $\lambda\sim 6-90$ cm with magnitudes $\gsim 0.1$Jy by the
ground-based radio telescope facilities such as those at Effelsberg,
Very Large Array, Westerbork Synthesis Radio Telescope, Owen's Valley
Radio Observatory, and Australia Telescope Compact Array. The optimal 
condition for a detection can be derived in advance by combining the 
information of locations of the Earth and the Jupiter relative to 
the Sun and the knowledge of low-latitude solar X-ray coronal holes 
where fast solar winds emanate. As the Sun rotates with an 
equatorial period of $\sim 25-26$ days, fast solar wind streams 
recur in the interplanetary space (Lou 1994, 1996).
Meanwhile, it is crucial to establish the correlation of QP-40 
pulsations of polar X-ray hot spots with arrivals of high-speed 
solar winds using the Chandra HRC. As Jovian auroral ovals mark 
the boundary zones separating the closed $\vec B$ of IRB and the 
open polar $\vec B$, it is inevitable that far-UV auroral ovals 
may pulsate with a QP-40 period upon arrivals of high-speed solar 
winds; this prediction can be tested by Hubble Space 
Telescope imaging spectrograph (HST-STIS) observations.
\footnote{Grodent, et al. (2001) indicated in their AGU Spring Meeting 
abstract \#P52A-09 that the sampling of the HST-STIS at the time did 
not permit them to highlight a forty-minute oscillation in the 
corresponding ultraviolet light curve. }

\subsection{Coincidence with a High-Speed Solar Wind}

For the Chandra HRC 10-hour observation for X-ray QP-40 pulsations 
of Jupiter's polar hot spot on 18 December 2000 from 10-20 UT 
(Gladstone et al. 2002), we show that this observation run 
happened to coincide with an arrival of high-speed wind at 
Jupiter. We obtained pertinent information from the website 
ssd.jpl.nasa.gov/
under Mean Orbital Elements for both the Earth and Jupiter,
and from the websites
www.sec.noaa.gov/ace/ACErtsw$\_$home.html and
www2.crl.go.jp/uk/uk223/service/arc/
for the archival data of real-time solar wind properties 
measured by the ACE spacecraft.
On 12:00 noon of 1 January 2000,
the Earth longitude L$_{E}=100.46435$ deg and the Jupiter
longitude L$_{J}=34.40438$ deg, respectively.
The mean angular rate of the Earth is $129597740.63''$/100yr=
$3548.1928''$/day=0.9856091 deg/day and the mean angular
rate of the Jupiter is $10925078.35''$/100yr=$299.11234''$/day
=0.08308676 deg/day. On 12:00 noon of 18 December 2000,
L$_{E}=87.398753$ deg and L$_{J}=63.65092$ deg. A high-speed 
wind of $\sim 700\hbox{ km s}^{-1}$ present at Jupiter on 
12:00 noon of 18 December 2000 should imply the leading edge of 
a low-latitude coronal hole at longitude L$_{H}=62.65$ deg on 
2:24 AM of 6 December 2000. As the equatorial angular rotation
rate of the Sun is $\sim 27$ days, 
a high-speed wind of $\sim 700\hbox{ km s}^{-1}$ is 
estimated to reach the Earth after 3:00 PM of 9 December 2000. 
According to the ACE data for the real-time solar wind speed 
$U$ in unit of $\hbox{km s}^{-1}$, we have
$390\lsim U\lsim 450$,
$550\lsim U\lsim 650$,
$650\lsim U\lsim 700$,
$700\gsim U\gsim 600$,
$600\gsim U\gsim 540$,
$550\gsim U\gsim 430$,
$440\gsim U\gsim 400$,
$U\sim 390$,
$380\gsim U\gsim 320$,
and $U\sim 330$ on $7-16$ December 2000, respectively. So a 
high-speed solar wind of $\sim 700\hbox{ km s}^{-1}$ did indeed 
arrive the Earth on 9 and 10 December 2000 by the ACE data.

As the polar X-ray hot spot is near the IRB, this observed result 
is consistent with the scenario that QP-40 IRB oscillations are 
excited more prominently at the arrival of high-speed solar wind. 
We need more Chandra HRC observations and correlation studies with 
the ACE solar wind speed data to firmly establish this important 
revelation. As the Jovian auroral ovals separate closed magnetic 
fields of the IRB from open magnetic fields of the pole, it is 
natural to expect that the far UV auroral ovals around the Jovian 
magnetic poles should also oscillate with a QP-40 period by the 
HST-STIS observations during that period of time.

\section{Summary and Conclusions}

Based on empirical clues, intuitive considerations, and 
theoretical analysis, it is physically plausible and 
appealing that polar QP-40 bursts of relativistic 
electrons and radio emissions involve global QP-40 IRB 
oscillations of Jupiter. With this scenario in mind, we 
summarize below key results and testable predictions.

\noindent
(1) QP-40 burst electrons, protons, and alpha particles with
relativistic energies most likely originated from the Jovian 
IRB (Lou 2001) where relativistic electrons with $\gamma$ as 
high as $\sim 100-200$ are known to exist (Bolton et al. 2002). 
They escape from the Jovian circumpolar region and are modulated 
by global QP-40 IRB oscillations. Some of escaping relativistic 
electrons may attain large $\gamma$ as indicated.

\noindent
(2) As estimated here, the predicted QP-40 brightness 
variations of the IRB should be observable by existing ground-based 
radio telescopes. It is extremely important to verify the expected 
correlation of such QP-40 brightness variations with arrivals of 
high-speed solar winds. Under favorable situations, higher harmonics 
of QP oscillations with shorter periods (Lou 2001) might also be 
detectable.

\noindent
(3) The QP-40 IRB oscillations are most likely excited and 
sustained by the combined effects of solar wind speed 
variations, of short-term intermittent wind speed 
variations during either fast or slow wind phase, and of 
angular momentum conservation of Jovian magnetospheric plasma.

\noindent
(4) Assuming a similar level of inhomogeneities for both north 
and south polar magnetic fields, we predict QP-40 bursts of 
relativistic electrons of the IRB from Jupiter's north polar 
region. Moreover, such QP-40 north polar activities are expected 
to correlate well with arrivals of high-speed solar winds at 
Jupiter. In coordination with spacecraft observations of solar 
wind properties, another spacecraft needs to be launched to 
enter Jovian magnetosphere and to probe the north polar region 
in order to test this prediction.

\noindent
(5) At $30-40R_J$ away from the Jupiter, synchrotron emissions 
with small pitch angles $\alpha$ are stronger than curvature 
emissions in QP-40 radio bursts associated with relativistic 
electrons streaming along south polar magnetic field lines of 
Jupiter. The observed predominance of right-handed polarization 
of radio burst emissions is consistent with the gyration sense 
of outstreaming electrons in the south polar magnetic field 
pointing towards Jupiter. North polar QP-40 radio bursts 
associated with north polar QP-40 bursts of relativistic 
electrons are expected and should be detectable by Ulysses, 
now approaching the Jupiter 
in the northern heliosphere. Because Jupiter's north polar 
magnetic field points outward, we thus predict the 
polarization of north polar QP-40 radio bursts to be 
predominantly left-handed. These predictions can be tested 
by Ulysses observations in the near future (year 2003-2004).

\noindent
(6) Based on one case of coincidence of QP-40 pulsations of north 
polar X-ray hot spots with an arrival of high-speed solar wind 
at Jupiter and our global QP-40 IRB oscillation scenario, we would
like to emphasize the importance of empirically establishing 
this correlation through more coordinated spacecraft observations. 

\noindent
(7) Latitudewise, the Jovian auroral ovals lie between the IRB 
and the polar X-ray hot spots. Through magnetospheric wave
transmissions of IRB oscillations to neighboring polar magnetic 
fields, we naturally expect QP-40 pulsations of Jovian aurora 
ovals in correlation with arrivals of high-speed solar winds
at Jupiter. For example, this prediction can be tested by 
well-prepared HST-STIS observations of FUV auroral ovals with 
appropriate sampling rate.

By this Letter, we hope to stimulate more observational and
theoretical studies on QP-40 polar activities of Jupiter 
and to identify the physical cause of QP-40 phenomena.

\section*{Acknowledgments}

This research has been supported in part 
by the ASCI Center for Astrophysical Thermonuclear Flashes 
at the University of Chicago under Department of Energy 
contract B341495,
by the Special Funds for Major State Basic Science
Research Projects of China, by the Collaborative
Research Fund from the NSF
of China (NSFC) for Young Outstanding Overseas Chinese
Scholars (NSFC 10028306) at the National Astronomical
Observatory, Chinese Academy of Sciences, and by the
Yangtze Endowment from the Ministry of Education through
the Tsinghua University.
Affliated institutions share this contribution.


\clearpage







\clearpage




\label{lastpage} 

\end{document}